\documentclass[twocolumn,aps,showpacs,prl,amsmath,amssymb,floatfix,superscriptaddress]{revtex4}
\usepackage{color}
\usepackage{graphicx}
\usepackage{dcolumn}
\usepackage{bm}
\usepackage{array}
\usepackage{float}
\usepackage{supertabular}
\usepackage{longtable}
\usepackage{mathrsfs}
\usepackage{txfonts}
\usepackage{wasysym}

\begin{document}

\title{Quantum Hall Ice}

\author{Gia-Wei Chern}

\affiliation{
Theoretical Division, T-4 and CNLS, Los Alamos National Laboratory, Los Alamos, NM 87545, USA}  
 \author{Armin Rahmani}
\affiliation{
Theoretical Division, T-4 and CNLS, Los Alamos National Laboratory, Los Alamos, NM 87545, USA} 

\author{Ivar Martin}
\affiliation{
Theoretical Division, T-4 and CNLS, Los Alamos National Laboratory, Los Alamos, NM 87545, USA} 
\author{Cristian D. Batista}
\affiliation{
Theoretical Division, T-4 and CNLS, Los Alamos National Laboratory, Los Alamos, NM 87545, USA} 

\date{\today}

\begin{abstract}

We show that the chiral kagome ice manifold exhibits an anomalous integer quantum Hall effect (IQHE) when coupled to itinerant electrons. Although electron-mediated interactions select a magnetically ordered ground state,  the full ice manifold can coexist with the IQHE over a range  of finite temperatures. The degenerate ice states provide a natural realization of power-law correlated flux disorder, for which the spectral gap of the system remains robust.  The quantized  (up to exponentially small finite-temperature corrections) Hall conductance persists over a wide range of electron densities due to the disorder-induced localization of electronic states.

\end{abstract}
 
\maketitle

The integer quantum Hall effect (IQHE) is  characterized by a topological invariant known as the first Chern number~\cite{Thouless1982}.
Besides the well-known two-dimensional (2D) electron gas in a magnetic field, quantum Hall effect can also emerge spontaneously from the interplay of itinerant electrons and local magnetic moments in the absence of an external magnetic field~\cite{Haldane1988,Taguchi2001}. The origin of this phenomenon lies in the Berry phases imparted to the electrons by magnetic textures with finite scalar spin chirality $\chi_{ijk} =\mathbf S_i\cdot\mathbf S_j \times \mathbf S_k$.
Such quantized Hall conductance can certainly be generated through long-range noncoplanar magnetic  ordering: $\langle {\mathbf S_j}\rangle,  \langle \chi_{ijk} \rangle \neq 0$~\cite{Ohgushi2000,Shindou2001,Akagi2010,Martin2008,Kato2010,Li2012,Yu2012,Venderbos2012}.
  Notably, the scalar spin chirality  can exist even without long-range magnetic order: $\langle {\mathbf S_j} \rangle = 0$ and $ \langle \chi_{ijk} \rangle  \neq 0$~\cite{Martin2008,Kato2010}. Indeed, for 2D systems, the chiral ordered phase  can persist at finite temperatures (it only breaks a discrete $Z_2$ symmetry), while magnetic order is destroyed by thermal fluctuations.~\cite{Mermin66}

The stability of 
the Hall conductance in such itinerant magnets is due to a robust locally ordered non-coplanar structure. While spin waves destroy the long-range magnetic order at finite temperature ($T$), long-wavelength distortions of the spin texture do not change the Berry flux pattern~\cite{Martin2008}. A quantized Hall liquid, however, can also be stabilized in a state with strongly disordered Berry fluxes, as we demonstrate for a geometrically frustrated  itinerant magnet. Up to exponentially small finite-$T$ corrections, an IQHE can coexist with an extensively degenerate ice manifold of magnetic local moments in a kagome lattice. This ``quantum Hall ice'' phase is a proof of principle for a new state of matter: an integer quantum Hall liquid coexisting with a classical spin ice. This work is partly motivated by recent experiments on a metallic spin-ice compound ${\rm Pr_2Ir_2O_7}$, which shows an anomalous Hall effect in the absence of any magnetic ordering~\cite{Nakatsuji2006,Machida2007, Machida2009}. 
\begin{figure}
 \includegraphics[width =0.95\columnwidth]{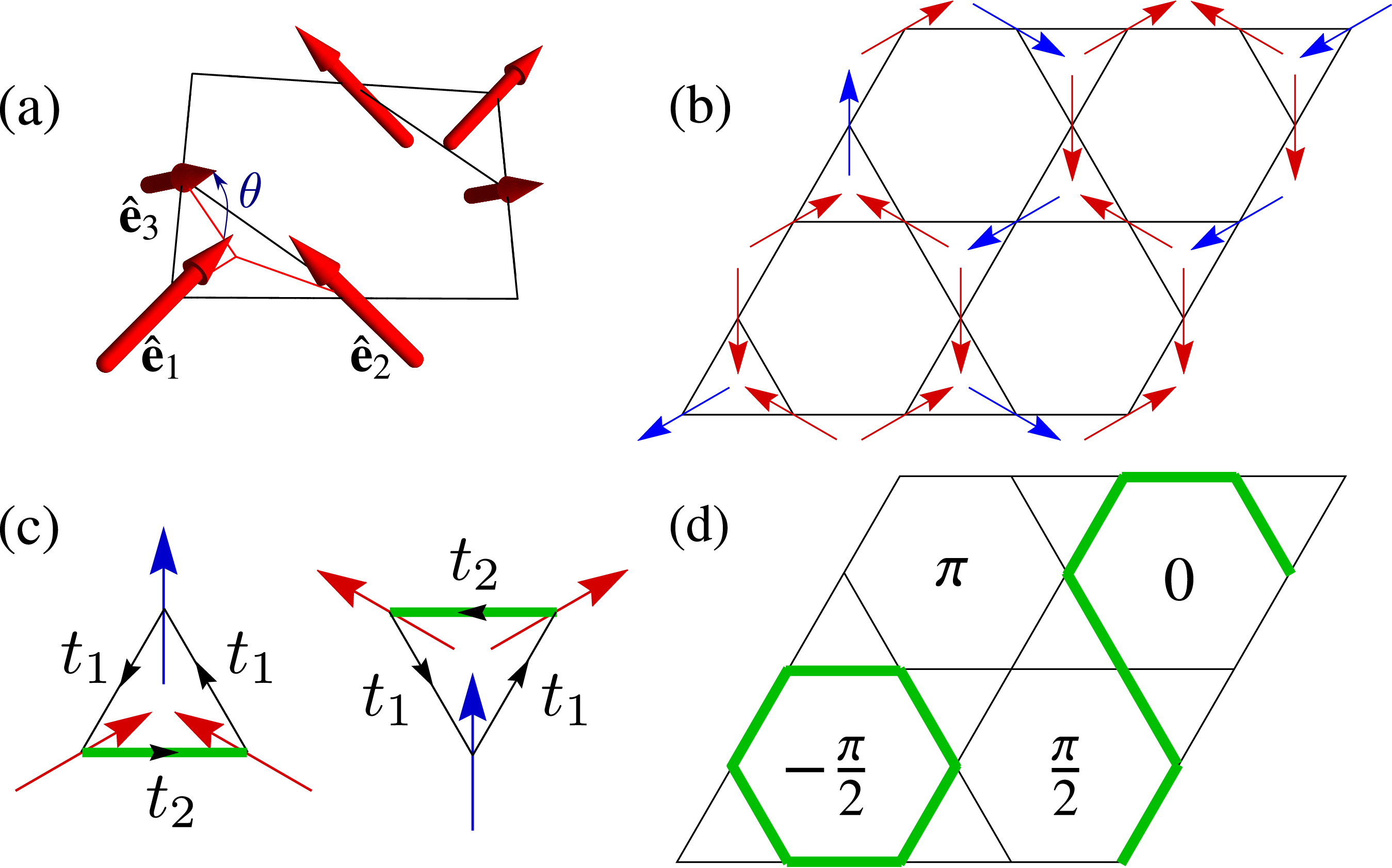}
 \caption{(a) The kagome lattice and the Ising axes $\hat{\bf e}_i$ for local magnetic moments ${\bf S}_i=\sigma_i \:\hat{\bf e}_i$. (b) The projection of ${\bf S}_i$ on the plane of the lattice for a random chiral kagome-ice configuration. Spins with $\sigma=+1$ ($\sigma=-1$) are shown in red (blue). (c) Two types of hopping amplitudes in a large-$J$ spinless effective model. (d) The fluxes in the effective model through the hexagonal plaquettes for the ice configuration of panel (b) at the special canting angle $\theta^*$.  }
 \label{fig:1}
\end{figure}

We show here that itinerant magnets with geometrical frustration provide experimentally relevant examples of correlated non-Gaussian disorder, which could have profound effects on the electronic states (such peculiar spin correlations, e.g., can explain the  resistivity minimum of metallic spin ice~\cite{Udagawa2012,Chern2012}). Despite great theoretical interest in the effects of correlated disorder on electron transport, experimental realizations have remained elusive~\cite{Ziman1961,Fisher1968,Sandler2004,Huckestein1995}. Moreover, non-Gaussian disorder is expected to give rise to novel phenomena~\cite{Javan-Mard2014} but rarely appears in nature (disorder from impurities in general have an uncorrelated Gaussian distribution).


One special feature of the peculiar power-law correlated flux disorder, originating from the ice rules, is the robustness of the spectral gap. Although amorphous solids are known to exhibit spectral gaps for strong disorder, survival of a spectral gap is unprecedented for strongly disordered IQHLs~\cite{Yang1996}. Another interesting feature concerns the properties of the magnetic monopoles (plaquettes that violate ice rules). Topological defects in itinerant magnets are known to exhibit unusual phenomena such as charge fractionalization~\cite{Muniz2012}. Here we demonstrate that pinned magnetic monopoles~\cite{Ryzhkin2005, Castelnovo2008} in the quantum Hall ice induce a fluctuating electric dipole in the charge density of the itinerant electrons.

We consider a kagome-lattice model in which Ising-like spins are subject to local constraints resembling the Bernal-Fowler ice rules~\cite{Bramwell2001}. This so-called ``kagome ice"~\cite{Wills2002} is an easy-axis ferromagnet with spins sitting on a two-dimensional network of corner-sharing triangles (Fig.~\ref{fig:1}). The projections of the local easy axes $\hat{\mathbf e}_i$ on the kagome plane form a 120-degree ordering, while the axes are canted with respect to this plane by an angle $\theta$. The spin direction  is specified by a set of Ising variables $\sigma_i$ as $\mathbf S_i = \sigma_i \,\hat{\mathbf e}_i$. The magnetic charges (in natural units) for every up and down triangles are $Q_\bigtriangleup=-\sum_{i \in\bigtriangleup}\sigma_i$ and  $Q_\bigtriangledown=+\sum_{i\in\bigtriangledown}\sigma_i$. The nearest-neighbor ferromagnetic exchange between spins $\mathbf S_i$ can be recast into $\sum_{\alpha} Q_{\alpha}^2$~\cite{Wills2002,Chern2011}, which penalizes triangles with magnetic charge $\pm 3$. It is thus energetically favorable for each triangle to have magnetic charges $\pm 1$. This implies the constraint that every triangle has either two incoming and one outgoing spins or vice versa.


 A subset of this kagome-ice manifold, known as the charge-ordered or \textit{chiral} kagome ice~\cite{Chern2011,Moller2009}, has a further constraint that spins in every up (down) triangle must be 2-in-1-out (1-in-2-out), i.e. all $Q_\bigtriangleup=-1$ and all $Q_\bigtriangledown=+1$~\cite{Chern2011,Moller2009}. Such configurations may be stabilized by two-body interactions of the form $Q_\bigtriangleup Q_\bigtriangledown$, long-range dipolar interactions~\cite{Chern2011,Moller2009}, or alternatively in spin-ice pyrochlores subject to a magnetic field in the [111] direction~\cite{Matsuhira2002,Udagawa2002} (in the presence of itinerant electrons, the 3D pyrochlore lattice may be approximated by decoupled 2D kagome layers if the inter-layer hopping is weak). Furthermore, it may be possible to stabilize chiral kagome ice through the electron-mediated interactions themselves.
 
  Although there is no long-range order in the Ising variables $\sigma_i$, the chiral kagome ice does have an overall magnetization in the out-of-plane direction,  and our results do not directly explain the experiments on ${\rm Pr_2Ir_2O_7}$, where the anomalous quantum Hall effect seems to persist even in the absence of net magnetization. Motivated by ${\rm Pr_2Ir_2O_7}$, however, the chiral itinerant kagome ice in a magnetic field was studied recently in the limit of low densities and small Kondo coupling, where an anomalous (nonquantized) Hall response was obtained~\cite{Udagawa2013}. Here we focus on the opposite limit of large Kondo coupling and finite densities in a broad range close $1/3$ filling and find a quantized response. 
 
%

We now introduce the itinerant electrons, which are coupled to the above Ising-like moments on the kagome lattice via an exchange coupling $J$. The electronic part of the Hamiltonian is then given by
\begin{equation}\label{eq:kondo}
H=t\sum_{\alpha\langle ij\rangle}\left(c^\dagger_{i \alpha} c^{\;}_{j \alpha}+{\rm H.c.}\right)+J\sum_{i \alpha \beta} {\bf S}_i \cdot c^\dagger_{i \alpha}{\boldsymbol \sigma}^{\;}_{\alpha\beta} c^{\;}_{i \beta},
\end{equation}
where $c_{i \alpha}$ is a fermionic annihilation operator on site $i$ and spin $\alpha$ and $t$ is the hopping amplitude.
If the classical energy scales are much larger than the electron-mediated spin-spin interactions, the coupling to itinerant electrons does not bring the system out of the ice manifold. Before a discussion of the energetics, we examine the fate of the electronic state for each chiral kagome-ice configuration of local moments ${\bf S}_i$.


\begin{figure}
\includegraphics[width =0.97\columnwidth]{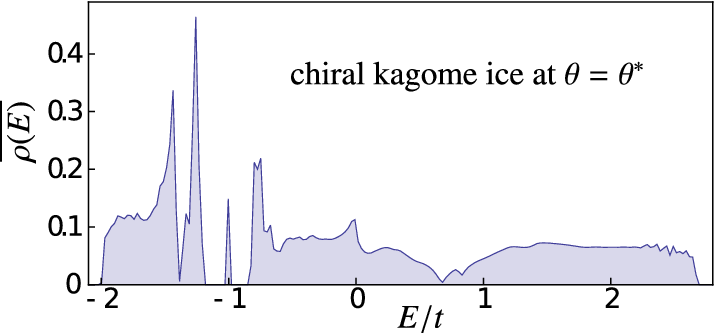}
\caption{(Color online) The disorder-averaged density of states $\overline{\rho(E)}$ for chiral kagome ice at $\theta=\theta^*$. }
\label{fig:3}
\end {figure}

To examine the intrinsic topological properties of the electrons in this ice manifold, 
we further simplify the problem by first considering the strong coupling limit. 
In the $|J|\gg |t|$ limit, the electrons align themselves with the local moments and the effective hopping amplitude between two sites with local moments ${\bf S}_i$ and ${\bf S}_j$ becomes $t\langle \chi_i |\chi_j \rangle$, where $|\chi_i\rangle$ is the spinor eigenstate of $\mathbf S_i\cdot\bm\sigma_{\alpha\beta}$.
As shown in Fig.~\ref{fig:1}(c), there are two distinct hopping constants:
$t_1=\cos {\pi \over 6} e^{-i{\pi \over 6}}\cos \theta$ and $t_2=\sin {\pi \over 6} e^{i{\pi \over 3}}+\cos {\pi \over 6} e^{-i{\pi \over 6}}\sin \theta$, for opposite and same Ising spins on the bond, respectively.
The gauge-invariant fluxes, which determine the electronic properties,  can be obtained from the above (gauge-dependent) amplitudes. These fluxes are equal in all up and down triangular plaquettes:
\begin{equation}\label{eq:flux_trig}
\Phi_\bigtriangleup=\Phi_\bigtriangledown=2\phi_1+\phi_2,\qquad \phi_i\equiv {\rm arg}(t_i).
\end{equation}
The fluxes in the hexagonal plaquettes of a generic chiral ice state, on the other hand, depend on the ice configuration, giving rise to a model of flux disorder.
These hexagonal fluxes are not uncorrelated. To see this, let us write the flux in a hexagon in terms of the six Ising variables around a hexagon as follows:
\begin{align} \label{eq:flux_hexagon}
\Phi_{\hexagon}= -6\phi_1 + (\phi_1-\phi_2) \textstyle\sum_{i \in \hexagon}\sigma_{i} ,
\end{align}
where $\sum_{i \in \hexagon}\sigma_{i}$ can take four distinct values: 0, 2, 4, and 6.  Due to a mapping of the chiral ice manifold to a dimer model on the honeycomb lattice~\cite{Udagawa2002,Moessner2003}, the Ising spins and consequently the fluxes above exhibit power-law correlations 
\begin{equation}
\label{eq:flux_corr}
\overline{\Phi_{\hexagon}({\bf  r})\Phi_{\hexagon}({0})}-\left(\overline{ \Phi_{\hexagon}({\bf  r})}\right)^2 \sim1/r^{2},
\end{equation}
where the ``overline'' indicates an average over chiral kagome ice configurations  (see the Supplemental Material for details). Note that the average flux in a hexagonal plaquette $\langle \Phi_{\hexagon}({\bf  r})\rangle=-4 \phi_1-2\phi_2$, is generically nonzero (the total flux through all triangles and hexagons vanishes). The same mapping to dimers also suggests that the flux disorder above is non-Gaussian.


The magnitude of the hopping amplitudes can, in general, take two different values $|t_1|$ and $|t_2|$. At a special canting angle $\theta=\theta^*={1\over 2}\arccos\left({1\over 3}\right)$, which we mostly focus on in the present paper, we have $|t_1|=|t_2|={t\over \sqrt{2}}\equiv \tilde{t}$. For this special $\theta$, the phases of the hopping amplitudes are given by $\phi_1=-{\pi \over 6}$ and $\phi_2={\pi \over 12}$, which according to Eqs.~\eqref{eq:flux_trig} and \eqref{eq:flux_hexagon} leads to a flux $-{\pi \over 4}$ in every triangular plaquette, and four different fluxes in the hexagonal ones, as shown in Fig.~\ref{fig:1}(d).

In an ordered $q=0$ configuration  [see Fig.~\ref{fig:1}(a)], which belongs to the chiral ice manifold, the flux in all hexagons is equal to $-2\Phi_\bigtriangleup$. The tight-binding Hamiltonian is known to exhibit a spectral gap and an IQHE at $1/3$ and $2/3$ filling fractions for this orderd state~\cite{Ohgushi2000}. The topological origin of the quantum Hall effect in the $q=0$ state can be understood by considering its band structure in the limits of $\theta = 0$ and $\theta={\pi \over 2}$.
The fluxes vanish in all plaquettes at $\theta=0$ but the tight-binding spectrum has pairs of Dirac points at  1/3 and 2/3 filling fractions. Non-zero fluxes resulting from canted spins gap out the Dirac points and lead to nontrivial band Chern numbers~\cite{Ohgushi2000,Haldane1988}. In the $\theta={\pi \over 2}$ limit, on the other hand, we obtain an array of one-dimensional chains. Moving away from $\theta={\pi \over 2}$ results in coupling these noninteracting Luttinger liquids (in the presence of time-reversal-symmetry-breaking fluxes), which also leads to quantum Hall effect~\cite{Sondhi2001, Kane2002}.


For random chiral ice states, the electrons experience 
flux disorder according to Eqs.~\eqref{eq:flux_hexagon} and \eqref{eq:flux_corr}. 
Generically, flux disorder should not differ from electrostatic potential disorder if there is a net flux through the system~\cite{Kalmeyer1993}. In case of the chiral kagome ice, the average flux vanishes, but since the system is characterized by a global quantum Hall response, one expects similar behavior to quantum Hall systems, which emerge in the presence of a net magnetic field.
 Generically, strong disorder closes the spectral gap in integer quantum Hall liquids (see, e.g., Ref.~\cite{Yang1996}) but the quantum Hall effect nevertheless persists at $T=0$. This can be understood in terms of a critical point (localization-delocalization transition) between two insulating states with extended states appearing only at a single critical energy $E_c$~\cite{Trugman1983,Chalker1988,Huckestein1990,Huo1992}. For the peculiar power-law correlated disorder considered here, we find that not only does the  quantized Hall conductance remain robust, but also the spectral gap at 1/3 filling remarkably persists even in the presence of strong flux disorder.

\begin{figure}
 \includegraphics[width =1.\columnwidth]{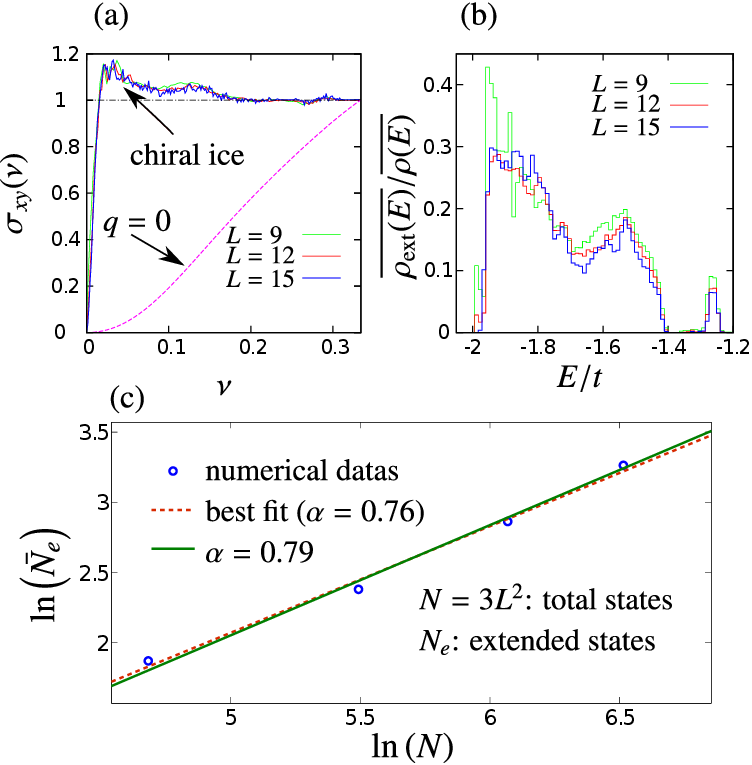}
 \caption{(Color online) (a) The disorder-averaged Hall conductance as a function of the Fermi energy. (b) The ratio of density of extended states to the total density of states. (c) The number of extended states as a function of the the system size. }
 \label{fig:4}
\end{figure}


Fig.~\ref{fig:3}(a) shows the disorder-averaged density of states obtained using loop-update Monte Carlo simulations at the special canting angle $\theta^*$. A spectral gap $\Delta \sim 0.2t$ can be clearly seen. Interestingly, we also observe a peak in the middle of the spectral gap. This peak corresponds to a flat band~\cite{Udagawa2013b} of localized states around hexagonal plaquettes with $ \Phi_{\hexagon} = -{\pi/ 2}$. 
%
%
We have checked that the spectral gap and a quantized Hall conductance in the kagome ice manifold remain robust for a broad range of canting angles, which include $\theta^*$ (see also Ref.~\cite{Ishizuka2013b} for a detailed study). The ice rules are indeed important for the robustness of the spectral gap. As shown in the Supplemental Material, both continuous and discrete uncorrelated random fluxes close the gap.


We computed the above-mentioned quantum Hall conductance $\sigma_{xy}$ explicitly for each ice configuration using the real-space version of the Kubo formula $\sigma_{xy}=\sum'_m \sigma^m_{xy}$, where $\sum'$ indicates summation over occupied levels $m$, and
\begin{equation}\label{eq:hall}
 \sigma^m_{xy}={2 e^2 \hbar \over A}\sum_{n\neq m} {\rm Im}\left[{\langle m|v_x|n\rangle \langle n|v_y|m\rangle\over (E_m-E_n)^2}\right],
\end{equation}
where $|n\rangle$ is a single-particle eigenstate with energy $E_n$, $A$ is the area of the system, and $v_i$ is the velocity operator in direction $i=x,y$.
We found that $\sigma_{xy}$ is indeed quantized at filling fractions $1/3$ and $2/3$ for all chiral kagome ice configurations.


Fig.~\ref{fig:4}(a) shows the $\sigma_{xy}$ as a function of the filling fraction $\nu$ for the $q=0$ state as well as the disorder-averaged $\sigma_{xy}$ for the full chiral kagome ice manifold. In order to suppress finite-size effects in the above calculation of $\sigma_{xy}(\nu)$, we average $\sigma_{xy}^m$ [Eq.~\eqref{eq:hall}] over various boundary phases, where a phase $0$ ($\pi$), e.g., corresponds to periodic (antiperiodic) boundary conditions. The Hall conductance rises much more quickly when increasing the density for chiral kagome ice than the clean $q=0$ ice state, and approaches its quantized value at a significantly smaller $\nu$. Similar to traditional quantum Hall states, where disorder leads to plateaus of transverse conductance as a function of the filling fraction, disorder stabilizes a $T=0$ quantized conductance over a wide range of filling fractions in chiral kagome ice.

The localization properties of the system can be explored in more detail by identifying localized and extended single-particle states. For each realization of the disorder, the angle-averaged $\langle \sigma_{xy}^m\rangle$ is an integral multiple of $e^2/h$~\cite{Thouless1984,Arovas1988,Yang1996,Yang1997,Yang1999,Sheng1995}. A state $|m\rangle$ with nonzero $\langle \sigma_{xy}^m \rangle \neq 0$ carries Hall current and is necessarily an extended state~\cite{Arovas1988}. This allows us to compute the average density of extended states. Fig.~\ref{fig:4}(b) shows the ratio of extended to total density of states. While the two higher energy peaks in Fig.~\ref{fig:4}(b)  decrease with increasing the system size, the lowest-energy peak persists and roughly coincides with the abrupt rise of $\overline{\sigma_{xy}}$ in Fig.~\ref{fig:4}(a). We estimate a critical energy $E_c=(1.9\pm 0.1)t$, close to the bottom of the band. This is in contrast to the traditional quantum Hall liquids, in which extended states appear in the middle of the broadened Landau level (the gap generally vanishes as Landau levels overlap). The proximity of  $E_c$  to the bottom of the band  explains why the Hall conductivity rises up quickly to its quantized value  in chiral kagome ice.

At the critical energy, $E_c$, the localization length is expected to diverge as $|E-E_c|^{- \nu}$ (the exponent $\nu$ not to be confused with the filling fraction). For Gaussian power-law correlated disorder with an exponent $2$, as in Eq.~\eqref{eq:flux_corr}, we expect this critical point to have an exponent $\nu=2.33$~\cite{Sandler2004,Huckestein1995}. We have checked the exponents for our non-Gaussian case by comparing the number of extended states $N_e$ (averaged over the kagome ice manifold) at a given system size with the total number of states $N$, which should go as $N^\alpha$, with $\alpha=1-1/2\nu$. As shown in Fig.~\ref{fig:4}(c), the results indicate that the localization exponent is approximately equal to $\nu=2.33$, as in the case of Gaussian disorder.

\begin{figure}
\includegraphics[width =0.97\columnwidth]{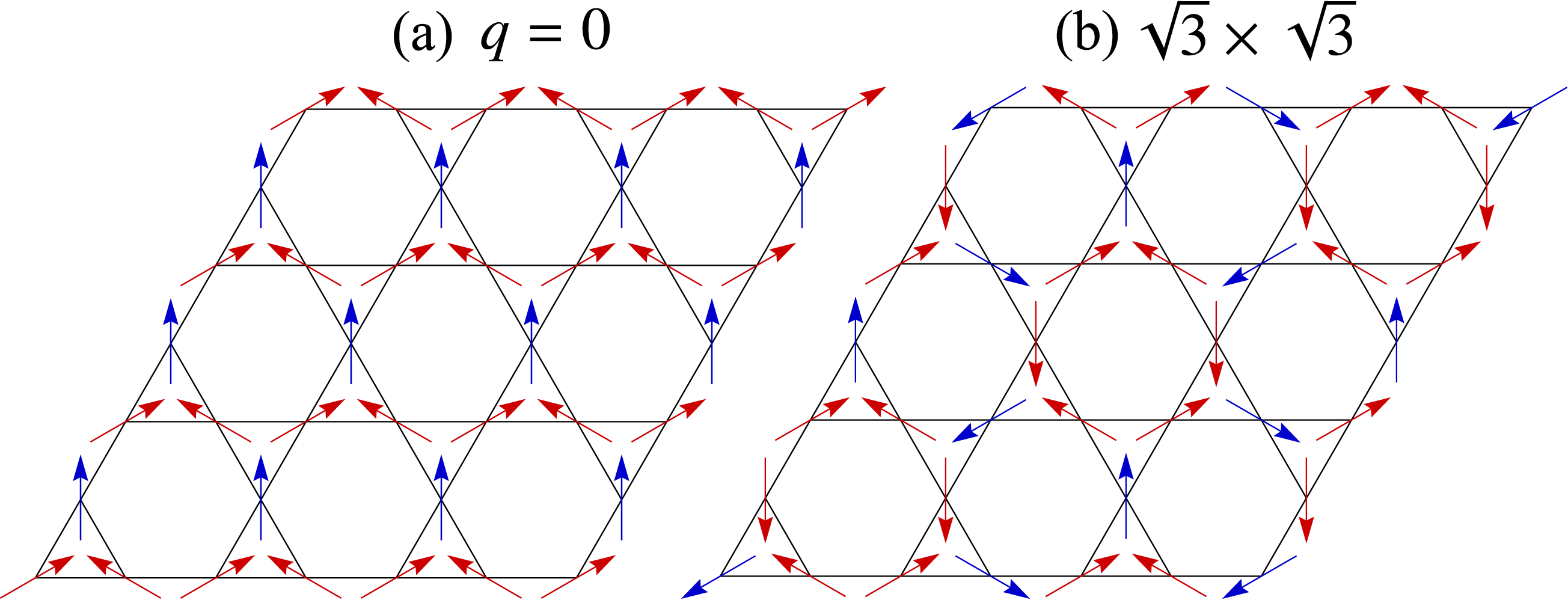}
\caption{(Color online) (a) The $q=0$ ice state. (b) The $\sqrt{3}\times\sqrt{3}$ ice state.}
\label{fig:2}
\end {figure}

We now address the energetics of the system in detail. So far, we assumed that the chiral kagome ice was stabilized by certain interactions between local moments and added the electrons in an \textit{ad hoc} manner. To study the energetic stability of this phase, we performed classical Monte Carlo simulations for large $J/t$, $\nu=1/3$ and $\theta=\theta^*$ and found that in fact at $T=0$, the $q=0$ [Fig.~\ref{fig:2}(a)] state has the lowest energy. However, this state is not a unique ground state: assuming periodic boundary conditions, we can transform the $q=0$ state to other ice states, with the same flux pattern, by flipping loops of spins. Similarly, the highest-energy state appears to be the $\sqrt{3}\times \sqrt{3}$  [Fig.~\ref{fig:2}(b)] state and related degenerate states, which give rise to the same flux pattern.

 As shown in Fig.~\ref{fig:2}, all chiral ice configurations give rise to a single-particle spectral gap $\Delta$ at $1/3$ filling fraction, which is typically an order of magnitude larger than the difference, $\delta$,  between the energy densities (per lattice site) of the  $q=0$  and $\sqrt{3}\times\sqrt{3}$ configurations. (At $\theta = \theta^*$ and $1/3$ filling, we have $\delta=0.02t$ and $\Delta=0.12t$.)  Since $\delta$ serves as a characteristic energy scale for electron-mediated spin-spin interactions, the energy difference between different chiral ice configurations is not resolved at temperatures $\delta \ll T \ll \min_{\{\sigma\}}\Delta (\{\sigma\})$, while the corrections to the quantized $\sigma_{xy}$ are exponentially small in the
 $\Delta/T$ ratio  over such range of temperatures.

After establishing the existence of the quantum Hall ice phase and studying its (bulk) topological and localization properties, we consider the interplay of magnetic monopoles with itinerant electrons in this system. Indeed, one of the most fascinating properties of spin ice is the emergent magnetic monopoles. These emergent excitations are defects violating the local ice constraints, which in the case of chiral kagome ice require $Q_\bigtriangleup=-1$ and $Q_\bigtriangledown=1$. Emergent magnetic monopoles in the chiral ice are triangles with $Q_\bigtriangleup=1$ and $Q_\bigtriangledown=-1$~\cite{Mengotti2011} (we neglect the higher-energy defects with $Q_{\bigtriangleup,\bigtriangledown}=\pm 3$).

In a given ice configuration, flipping an open string of head-to-tail spins can create two such triangles, which violate the ice rules. The defects in turn change the Berry flux pattern experienced by the electrons (see Supplemental Material for details). Interestingly, an incarnation of Laughlin's flux-insertion argument in this quantum Hall state predicts the formation electric dipoles  pointing from the center of the (monopole) triangle to the center of the neigboring hexagons. Detection of such dipoles in a given realization is difficult because the charge density is not generically uniform, and flipping a finite string creates charge density variations with a similar magnitude to the variations already present in a typical configuration.

\begin{figure}
\includegraphics[width=0.6\columnwidth]{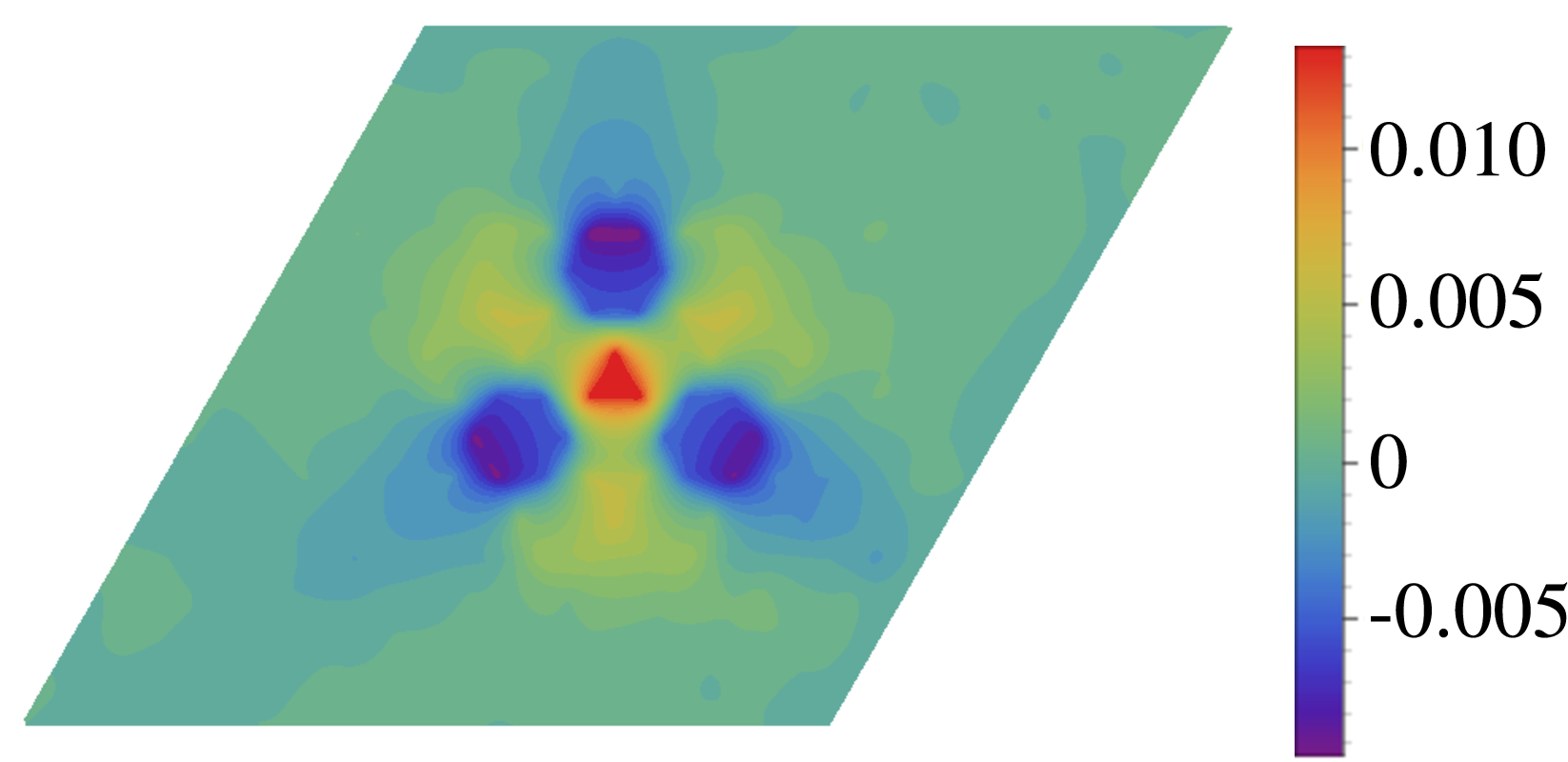}
\caption{\label{fig:monopole2}Numerically computed charge density in the vicinity of a pinned magnetic monopole, averaged over ice realizations with two monopole defects (the shown region has  24$\times$24 lattice spacings). }
\end{figure}

However, magnetic monopoles, which are in fact defined
for the full fluctuating spin ice manifold, can be pinned by impurities. For two such pinned monopoles, the charge density profile is uniform away from the defects after averaging over a time scale longer than the characteristic time scales of the spin dynamics. The remnants of the above-mentioned electric dipoles appear around each monopole as three electric dipoles with $C_3$ symmetry (see Fig.~\ref{fig:monopole2}).  


In summary, the magnetic exchange between conduction electrons and a frustrated set of local  Ising moments  can lead to a novel state matter, simultaneously characterized  by spin-ice local moment physics  and a quantized (with exponentially small finite-$T$ corrections) anomalous quantum Hall effect for itinerant electrons over a wide range of filling factors. This spontaneously broken symmetry state does not require of any external magnetic field or spin-orbit interaction. The critical correlations of spin ice produce a peculiar form of power-law correlated flux disorder that has nontrivial consequences on the spectrum and transport properties
of the conduction electrons. While previous studies have focused on the longitudinal conductivity of ``metallic-ice'' \cite{Udagawa2012,Chern2012}, we have shown  that {\it chiral } spin-ice can dramatically change the electronic state by inducing a robust Quantum Hall liquid (``Quantum Hall ice''). Moreover, the interplay of the electrons with magnetic monopole defects  may provide novel electronic signatures for detecting these monopoles.

\acknowledgements
We are grateful to C. Castelnovo, S. Trugman, and K. Yang for helpful discussions. This work was supported by the U.S. DOE under LANL/LDRD program (C.B., G.W.C., I.M., and A.R.) and a LANL Oppenheimer fellowship (G.W.C.). {Upon completion of this work, we became aware of an independent parallel work~\cite{Ishizuka2013}, in which the quantum Hall ice phase was also predicted in full agreement with our findings. Beyond this fundamental result, these two related works address different aspects of the problem (e.g., localization and topological defects in the present manuscript).}

\bibliography{ice}{}

\end{document}


\title{Quantum Hall Ice: Supplementary Materials}
\author{Gia-Wei Chern}

\affiliation{
Theoretical Division, T-4 and CNLS, Los Alamos National Laboratory, Los Alamos, NM 87545, USA}  
 \author{Armin Rahmani}
\affiliation{
Theoretical Division, T-4 and CNLS, Los Alamos National Laboratory, Los Alamos, NM 87545, USA} 

\author{Ivar Martin}
\affiliation{
Theoretical Division, T-4 and CNLS, Los Alamos National Laboratory, Los Alamos, NM 87545, USA} 
\author{Cristian D. Batista}
\affiliation{
Theoretical Division, T-4 and CNLS, Los Alamos National Laboratory, Los Alamos, NM 87545, USA} 

\date{\today}

\maketitle

\section{Random flux model}

The robustness of the spectral gap in the itinerant chiral kagome ice is unusual. As two distinctive characteristics of the fluxes in chiral kagome ice are discreteness and power-law correlations, it is illuminating to study random flux models~\cite{sheng95,yang97}, which do not possess one or both of these features. To this end, we consider the following spinless tight-binding Hamiltonian on the kagome lattice,
\begin{eqnarray}
	H = -t \sum_{\langle ij \rangle} \left(e^{i \phi_{ij}}\, c_i^{\dagger} c^{\;}_j + \mbox{h.c.}\right).
\end{eqnarray}
Here $t$ is a real-valued hopping constant, $c^\dagger_i$ ($c_i$) are the electron creation (annihilation) operators, $\langle ij \rangle$ runs over all nearest-neighbor bonds, and $\phi_{ij}$ are random variables satisfying the following constraints:
\begin{eqnarray}
	\label{eq:constraints}
	\sum_{\alpha \in \bigtriangleup} \phi_{\alpha} = \Phi, \qquad 	
	\sum_{\alpha \in \bigtriangledown} \phi_{\alpha} = \Phi,
\end{eqnarray}
where we set the constant flux $\Phi = -\pi/4$ as in the case of chiral kagome ice for $\theta=\theta^*$. We consider two different cases: in the first case, the phases $\phi_{ij}$ are continuous random variables subject to the Eq.~(\ref{eq:constraints}); see also the inset of Fig.~1(a). Practically, two of the phases $\phi_{ij}$ in a triangle are random variables uniformly distributed between 0 and $2\pi$, while the third one is determined by the above constraints. It is obvious that the fluxes on hexagons are continuous and uncorrelated. As shown in Fig.~1(a), the spectral gap completely closes in the disorder-averaged density of states.

In the second case, we retain the discrete nature of the random fluxes. As discussed in the main text, each triangle in chiral kagome ice has two (one) bonds with phase $\phi_1$ ($\phi_2$). Again we have $2\phi_1 + \phi_2 = \Phi$; see the inset of Fig.~1(b). However, there are additional constraints imposed by the charge-order in chiral kagome ice, which make it a critical phase. We can take away these additional constraints, and randomly place one bond with phase $\phi_2$ on each triangle (on phase $\phi_1$ on the other two bonds). Such discrete model is very close to chiral kagome ice except now the fluxes on far away hexagons are uncorrelated. 
\begin{figure}
\includegraphics[width=0.85\columnwidth]{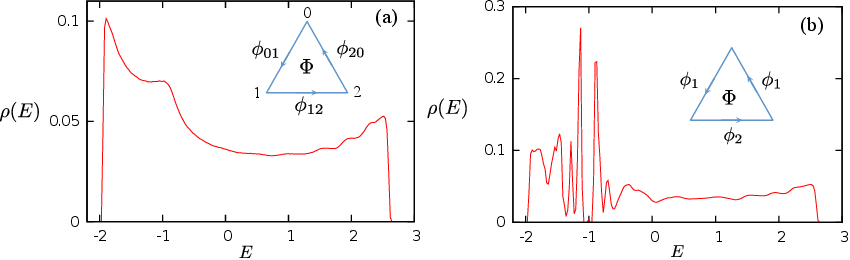}
\caption{Density of states for (a) the continuous and (b) discrete random flux models. The insets show the corresponding constraints for the random phases $\phi_{ij}$ on an elementary triangle of the kagome lattice. Here we choose $\Phi = \pi/4$.}
\end{figure}

We show the density of states for such manifold (with $\phi_1$ and $\phi_2$ corresponding to the special angle $\theta^*$ discussed in the main text) in Fig.~1(b). We find that the localized states inside the gap spread out to other energies. The average density of states exhibits an exponentially small number of states in the energy range corresponding to the spectral gap of chiral kagome ice. Most individual configurations are still gapped. 
We thus conclude that flux correlations are necessary for a fully gapped spectrum.

\section{Power-law flux-flux correlation}

We first discuss the mapping to a dimer model on the honeycomb lattice:
connecting the centers of the  corner-sharing triangles in the original kagome lattice creates a dual honeycomb lattice. Placing a dimer between the centers of two neighboring triangles that share a minority spin $\sigma_i = -1$,
maps each chiral ice state a dimer covering on the dual honeycomb lattice (as each triangle has only one minority spin). The ensemble of all dimer coverings form a critical phase with dimer-dimer correlations decaying asymptotically as $1/r^{2}$. These dimer correlations readily translate to flux correlations disciussed in the main text. The mapping of the dimer model to a height model in turn indicates suggest that the dimer correlations do not satisfy the Wick's theorem and, consequently, the disorder must be non-Gaussian.

Here we present a numerical verification of the power-law flux correlations originating from the chiral ice rules~\cite{moller09,chern11,udagawa02,moessner03}.  
 As discussed in the main text, spin states in the chiral ice manifold should exhibit algebraic power-law behavior at long distances due to the mapping to a critical dimer model. Using nonlocal loop updates~\cite{chern11}, we generated random configurations in the chiral kagome ice manifold, and computed the average flux correlations. Fig.~2(a) shows the short-range hexagonal plaquette flux-flux correlation function: $C_{\Phi}(r) = \overline{ \delta\Phi(r)\, \delta\Phi(0) }$.  Here $\delta\Phi(\mathbf r) = \left(\Phi(\mathbf r) - \overline{ \Phi } \right)/\pi$,  and the average flux is $\overline{ \Phi }  = -\pi/2$. It exhibits a period-3 damped oscillation, indicating a local $\sqrt{3}\times\sqrt{3}$ spin order.  This is consistent with the fact that the $\sqrt{3}\times\sqrt{3}$ state (corresponds to a flat height phase) has the maximum number of flippable hexagonal loops and is favored entropically.  At large distances, our simulations shown in Fig.~2(b) find a power-law correlation $C_{\Phi}(L/2) \sim 1/L^2$, consistent with the exact solutions~\cite{moessner03}.
\begin{figure}
\includegraphics[width=0.85\columnwidth]{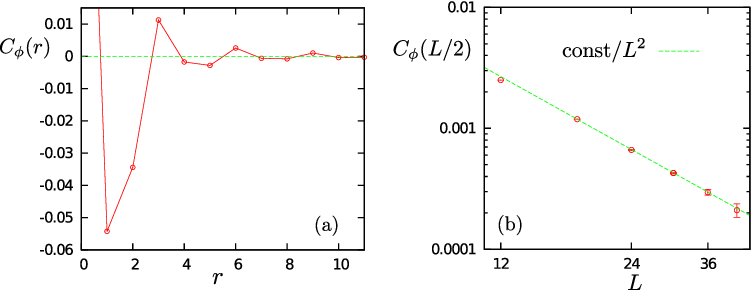}
\caption{(a) Hexagonal plaquette flux-flux correlation function: $C_{\Phi}(r) = \overline {\delta\Phi(r)\, \delta\Phi(0) }$. For any given ice states of the chiral kagome ice manifold (considering only the case of critical angle $\theta = \arctan\sqrt{2}$, each triangle (both types) has exact flux $\Phi_{\bigtriangleup} = \Phi_{\bigtriangledown} = \pi/4$. On the other hand, the fluxes on the hexagonal loops can assume $\Phi_{\hexagon} = -\pi$, $-\pi/2$, 0, or $\pi/2$, with the constraint that the average flux is exactly $-\pi/2$.  (b) Finite-size scaling of the flux-flux correlation follows a power-law decay $C_{\Phi}(L/2) \sim 1/L^2$. The correlation function $C_{\Phi}(L/2)$ is obtained for a given size $L$ by measuring flux correlation between two hexagons separated by half the lattice size $L/2$ with periodic boundary conditions. }
\end{figure}

\section{Magnetic monopoles}
\begin{figure}
\includegraphics[width=0.75\columnwidth]{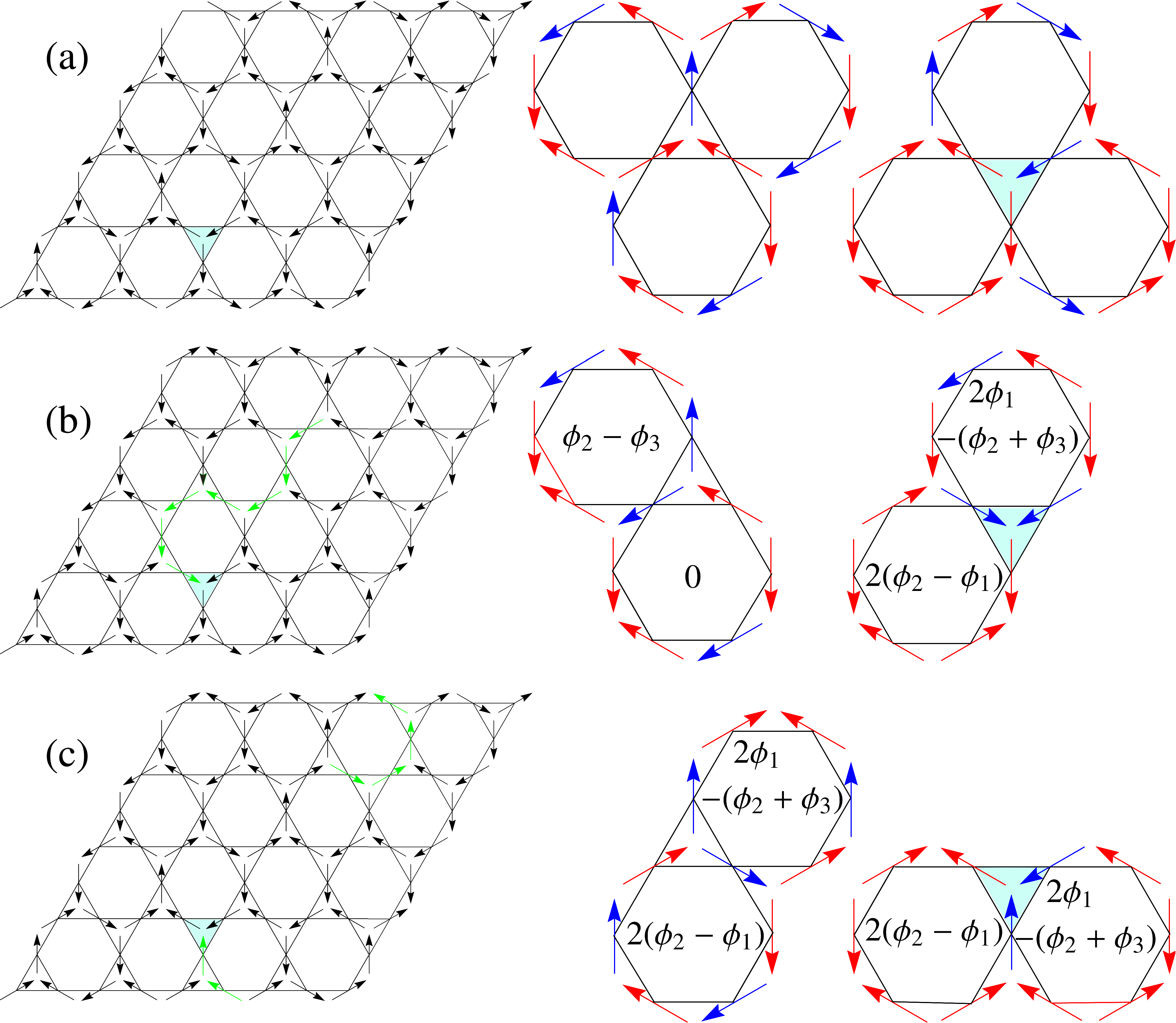}
\caption{\label{fig:monopole1}(a) A random chiral kagome ice configuration. Magnetic monopoles will be created in the two colored triangles by flipping a Dirac string connecting these triangles. The Ising spins in the vicinity of these triangles are shown in red ($\sigma=1$) and  blue ($\sigma=-1$) in a zoomed-in image. (b) and (c) The magnetic monopoles are created by two different Dirac strings (flipped spins with respect to the  (shown in green). Similarly the Ising spins in the monopole triangle and the hexagons that share the last flipped spin in the Dirac string are shown on the right-hand side. The additional fluxes in these hexagons [with respect to the bare configuration in panel (a)] are also shown. }
\end{figure}

Here we discuss the variation of the flux pattern in the presence of magnetic monopoles. In Fig.~\ref{fig:monopole1}(a), we show a random chiral kagome ice configuration. We will create two magnetic monopoles (triangles with magnetic charge $Q_{\bigtriangleup,\bigtriangledown}=\mp 1$, which violate the ice rules of the chiral manifold ) in the two colored triangles by flipping a Dirac string connecting the triangles. In Figs.~\ref{fig:monopole1}(b) and ~\ref{fig:monopole1}(c), we show two such configurations corresponding to two different Dirac strings. The presence of defects creates a third type of effective hopping amplitude. In the chiral ice manifold, when moving counterclockwise in a triangle, we had hopping $t_1$ ($t_2$) on a bond with two Ising spins that had different sign (were both positive). Now, we can have a third type of bond with two negative Ising spins, a hopping amplitude $t_3=\sin {\pi \over 6} e^{i{\pi \over 3}}-\cos {\pi \over 6} e^{-i{\pi \over 6}}\sin \theta$, and a corresponding phase $\phi_3\equiv {\rm arg} (t_3)$.

The flux in the two triangles that correspond to magnetic monopoles then changes from $2\phi_1+\phi_2$ to $2\phi_1+\phi_3$, indicating the insertion of an additional flux $\phi_3-\phi_2$. By examining the Ising variables around the neighboring hexagons, we can compute their flux variations. We find that the flux $-(\phi_3-\phi_2)$ is either inserted in one of the neighboring hexagons or distributed between two of them. If the monopoles are pinned, the system fluctuates between configurations with these two pinned monopoles. Using Laughlin's argument, we find from the flux pattern a $C_3$ symmetric charge distribution corresponding to three dipoles in the directions pointing from the triangular monopoles to the centers of their three neighboring hexagons (assuming that the measurement time is longer than than spin fluctuation time scales). This is indeed confirmed numerically as seen in Fig.~5 of the main text. Note that in a typical individual realization, the background charge density is strongly disordered and the monopoles can not be easily identified from the density profile. Nevertheless, the distinctive electronic signature above emerges in the time average of the charge density for \textit{pinned} magnetic monopoles.